\documentstyle[11pt,paspconf]{article}
\def\la{\mathrel{\mathpalette\fun <}}
\def\ga{\mathrel{\mathpalette\fun >}}

\def\fun#1#2{\lower0.837ex\vbox{\baselineskip0ex\lineskip0.209ex
  \ialign{$\mathsurround=0ex#1\hfil##\hfil$\crcr#2\crcr\sim\crcr}}}
\def\memsai{\ref@jnl{Mem.~Soc.~Astron.~Italiana}}

\def\msun{M_\odot}

\def\sles{\lower2pt\hbox{$\buildrel {\scriptstyle <}
   \over {\scriptstyle\sim}$}}
 
\def\sgreat{\lower2pt\hbox{$\buildrel {\scriptstyle >}
   \over {\scriptstyle\sim}$}}

\def\la{\mathrel{\mathpalette\fun <}}
\def\ga{\mathrel{\mathpalette\fun >}}

\def\msun{M_\odot}
\def\la{\mathrel{\mathpalette\fun <}}
\def\ga{\mathrel{\mathpalette\fun >}}

\begin{document}

\title{ Accretion Disks in Transient 
      Systems}


\author{ John K. Cannizzo\altaffilmark{1}}
 \altaffiltext{1}{to appear in The 13th
    North American Workshop on Cataclysmic Variables,
   ed. S. B. Howell, E. Kuulkers, \&
   C. Woodward (San Francisco: ASP)  }
\affil{Universities Space Research Association,
  NASA/GSFC/Laboratory for High Energy Astrophysics,
    Greenbelt, MD 20771 }





\begin{abstract}
I review recent advances in our understanding 
of accretion disks in transient systems
$-$ the dwarf novae and the soft X-ray transients.
The primary theme will be
the ongoing development
of theory in response
to the observations.
The accretion disk limit
cycle model
appears to provide
a unifying 
framework within which
we may begin to understand
what is seen in
different types of 
    interacting
binary stars,
and also
to constrain
parameters which enter
into the theory.
\end{abstract}




\section{Introduction}

The last 15 years have witnessed major advances
in our understanding of the steady
state and time dependent nature of
accretion disks
in dwarf novae (DNe) and soft X-ray transients (SXTs).
These systems consist
  of semi-detached
  binaries
    in which a Roche lobe
   filling K or M star
  transfers matter into an accretion
   disk surrounding a compact object.
  In the DNe the compact
object is a white dwarf (WD),
whereas in the SXTs
it is a neutron star (NS)
or  black hole (BH).
The   advances
   discussed in this work
     have come about largely because
of the development
of the accretion disk limit cycle mechanism
as a model to account for outbursts
in DNe and SXTs.
The outbursts one sees
in these systems typically
last for days
to weeks and recur on intervals
of weeks to years.
Such short time scales,
coupled with the relative
brightnesses of some of the nearer
systems,
make these objects
ideal laboratories
for studying the physics of accretion disks.

In this review I will touch briefly
on six aspects
of the limit cycle 
instability $-$
with an emphasis
on the connection
between observation and theory.
These areas are:

(i) the steady state physics underlying the instability,

(ii) the general time dependent evolution of disks
   subject to the limit cycle instability,

(iii) the ``problem'' of the delay of the UV flux with
respect to the optical during the onset of a fast-rise
outburst,

(iv) Warner's (1987) $M_V$(peak) versus $P_{\rm orbital}$
relation for DNe at maximum light,
and its relevance to the $M_V$ versus
$(L_X/L_{\rm Edd})^{1/2} P_{\rm orbital}^{2/3}$
relation for SXTs found by van Paradijs \&  McClintock
(1994),

(v) the implication of observed exponential
   decays of outbursts seen in dwarf novae
   and SXTs for the Shakura \& Sunyaev (1973)
   viscosity parameter $\alpha$, and

(vi) the effect of evaporation of material
   from the inner disk on the long term light
curves of the SXTs.

The work I will discuss makes
the fundamental assumption
that our steady state and time 
dependent modeling
has direct relevance for 
    the outbursts that
we observe in DNe and SXTs,
and that we can therefore work
backwards from the observations
to place limits on the magnitude
of the viscosity parameter $\alpha$.
In the early stages of this line of inquiry,
most modelers were content with simply
estimating how big $\alpha$ is.
More recently we are trying to
go one step further and place limits
on the functional dependence of $\alpha$
on local physical variables.
The work described by John Hawley
represents the more fundamental
effort in this regard as it
attempts to calculate the viscous
dissipation from first principles
(Hawley, Gammie, \& Balbus 1995, 1996).
The ultimate goal is for workers
in these two philosophically distinct
     camps to compare  results
and see how their findings  about
the viscosity $-$
usually expressed in terms of $\alpha$
for historical reasons $-$
agree or disagree.
Perhaps in another five to 10 years
workers 
will begin to be able
to cross-compare findings
between the two groups.

My personal strategy
in  carrying out the task
of going from observation
to theory is that I model
the brightest and
best studied objects.
This means that for DNe,
I take SS Cygni,
and for SXTs,  I take
A0620-00.
SS Cyg is the brightest DN
in the sky.
Its orbital period is $\sim6.6$ hr
(Joy 1956),
and it shows outbursts
of amplitude $\delta m_V\simeq3.3$
which last  for $\sim10-20$ days
and recur every $\sim2$ months
(Campbell 1934, Cannizzo \& Mattei 1992).
A0620-00
is a black hole candidate
with $P_{\rm orbital}\sim7.75$ hr
(McClintock \& Remillard 1986)
which had the brightest
X-ray outburst
seen in an SXT (Nova Mon 1975).
It is the only SXT
which lies within 1 kpc
of Earth (Chen, Shrader, \& Livio 1997).

\section{Background/Steady State Physics}

Although one can find in
the pre-1980's literature certain
statements which foresaw in some way
the possible operation of a disk instability
as the agent responsible
for DN outbursts,
one can trace the roots of modern day
research on the limit cycle instability
back to the Sixth North American
Workshop on  Cataclysmic
Variable Stars
and Low Mass X-ray Binaries
which was held
in the summer of 1981
in Santa Cruz.
At this meeting
Jim Pringle 
discussed in general terms
the relation between the
integrated viscosity $\nu\Sigma$
and the surface density
$\Sigma$ which an accretion
disk must possess
in order to obtain
limit cycle behavior.
The parameter $\nu$
is the ``kinematic viscosity coefficient'',
$\nu=2\alpha P/(3\Omega\rho)$,
and $\alpha$ is the Shakura \& Sunyaev (1973)
parametrization of the $\phi-r$
component
of the viscous
     stress tensor
expressed in terms of the local pressure $P$.
When $\nu\Sigma$  is  plotted
in terms of $\Sigma$,
one sees an ``S'' shaped curve.
In the audience at this meeting
were several accretion disk modelers,
notably Friedrich Meyer, Doug Lin,
and John Faulkner.
Meyer had just completed a paper
with his wife
    Emmi Meyer-Hofmeister
on the vertical structure
of accretion disks
which was submitted
the same month as the meeting
(and appeared early in 1982),
but nowhere in their paper had they
thought to plot their solutions
of the disk structure
as $\nu\Sigma$ versus $\Sigma$.
Traditionally,
one had always plotted
all physical
quantities versus radius.
Upon his return to Germany, Meyer
produced a  short paper with Meyer-Hofmeister
using some of the solutions
taken from their structure paper.
It was submitted to Astronomy \& Astrophysics
in October 1981,
and it appeared in December 1981.
Figure 2 in this
article shows a plot
of ${\dot M}$
(which is equivalent to $\nu\Sigma$,
and which also
   scales with
        $T_{\rm midplane}$ or $T_{\rm effective}$)
    versus $\Sigma$
   computed from 
   the vertical
   structure at one radius
and for one $\alpha$ value
(Meyer \& Meyer-Hofmeister 1981, 1982).
In this figure we see the hysteresis
between ${\dot M}$ and $\Sigma$.

Shortly thereafter
other workers,
 many  of whom were not
   at the Santa Cruz meeting,
   became
aware of the importance of
calculating  the vertical
structure of accretion disks,
and   numerous other
papers appeared showing the
S-curves for rates of accretion
giving disk temperatures
near $\sim10^4$ K,
the dividing line between
neutral and ionized hydrogen
(cf. Cannizzo 1993a for a complete
list of references).
The physics controlling the vertical structure
depends heavily on
    the opacity.
   For $T\ga10^4$ K,
the opacity is large because
of the ready availability
of free electrons
to contribute to bound-free and free-free
processes.
The opacity 
$\kappa\sim\rho^a T^b$
with $b\sim-2$ to $-3$.
For $T\la10^4$ K
the opacity falls steeply 
with decreasing temperature
      as electrons
and protons recombine.
For 6000 K $\la T \la$ 8000 K,
one has $b\sim10-20$,
and for $T\la$ 6000 K
the opacity is small.
The form of $\kappa$ at low
temperatures is complicated
due to its dependence on
a variety of molecules
(Cannizzo \& Wheeler 1984).
It is the large change
in both the amplitude
and functional form
of $\kappa(\rho,T)$
between neutral and ionized
gas which ultimately
leads to the shift in the ${\dot M}-\Sigma $
(or, equivalently, $T_{\rm eff}-\Sigma $)
relation between the lower and   upper
branches   of the S-curve.

The specific shape and scalings
of the local minima $\Sigma_{\rm min}$
and maxima $\Sigma_{\rm max}$
in $T_{\rm eff}(\Sigma)$
are dependent somewhat
on the assumed strength
of convection,
usually cast in terms
of $l/H_P$,
the ratio of the mixing length
to pressure scale height
(Mineshige \& Osaki 1983,
 Pojma\'nski 1986, 
Cannizzo 1992, Ludwig, Meyer-Hofmeister, \& Ritter 1994).
Broadly speaking,
$\Sigma_{\rm max}$
and $\Sigma_{\rm min}$
increase with $l/H_P$  $-$
    this effect being 
   more  pronounced
for smaller
    $\alpha$ values.
All studies
to date
  which integrate
the steady state vertical structure
   equations
assume standard stellar
mixing length
theory,
which is questionable
for accretion 
   disks because of the   strong shear.
Nevertheless,
it seems  difficult to imagine
that a more correct treatment
of convection
by researchers in the distant
future could
somehow eliminate
the S-curve
and produce a monotonic $T_{\rm eff}-\Sigma$
relation,
given that the opacity change
is more fundamental
than convection
to the overall $T_{\rm eff}-\Sigma$
morphology.
Indeed,
the omnipresence
of the hysteresis
in so many articles
by independent
groups
gives us confidence in its reality.

A separate issue is that of
viscous dissipation
and angular momentum transport
induced by convective
motions within the accretion
disk vertical structure.
John Hawley argued convincingly
that any such mechanism
related to turbulence
which couples linearly
to the shear
cannot transport
angular momentum
outward
as required
(cf. Balbus, Hawley, \& Stone 1996).
Many previous workers 
had treated  turbulent 
heat transport and angular momentum
transport
  in accretion disks
    collectively
   as ``passive contaminants''
  which have similar
  physical properties,
  in spite of much work to
suggest the contrary
(for a discussion
see Balbus \& Hawley 1997).
Calculations show only
a weak inward transport
of angular momentum
if  convection is the 
relevant agent
(Ryu \& Goodman 1992,
 Stone \& Balbus 1996).
Although a few early limit
cycle papers
tried to connect $\alpha$
to some function involving
the ratio of the convective
speed to the sound speed
(e.g., Smak 1982, Cannizzo \& Cameron 1988,
Duschl 1989),
no one currently active in the field continues
to advocate such a model.


%

\section{Time Dependent Behavior/Outburst Rise Time
   versus Triggering Location within the Disk}

Soon after the first
vertical structure
papers appeared,
workers began to consider
the global,
time-dependent evolution
of the disk in response to the
 S-curve relation which exists
at every  radius.
The instability was found to operate
in the following manner:
During the time corresponding 
to the quiescence
interval for DNe,
matter arriving at the outer
disk edge from the secondary star
piles up at large radii,
thereby augmenting
$\Sigma$.
When one plots $\Sigma$
as functions of $r$ and $t$,
one sees that,
in quiescence,
$\Sigma(r)$
is bounded by $\Sigma_{\rm min}(r)$
and $\Sigma_{\rm max}(r)$.
These two physical quantities
scale linearly with radius,
so the mass enclosed within a cylinder
of radius $r$ 
scales steeply with radius,
$M(r)\sim r^3$.
Thus most of the mass lies at large radii.
As mass continues to accumulate,
eventually we must have
$\Sigma(r_{\rm trigger})> \Sigma_{\rm max}(r_{\rm trigger})$
at some radius $r_{\rm trigger}$.
The radius of onset of the instability
$r_{\rm trigger}$
can lie anywhere in the disk.
As gas within this annulus becomes unstable,
it begins to heat quite vigorously,
and its movement
in $(T_{\rm eff},\Sigma)$ or $(T_{\rm mid},\Sigma)$
takes it from the lower to the upper
stable branch of the S-curve,
although often by a circuitous route
(Mineshige \& Osaki 1985,
Fig. 12).
   The viscosity coefficient
$\nu$ scales  as $\alpha T_{\rm midplane}$,
so the factor of $\sim100$
increase in $T_{\rm midplane}$
and thence $\nu$ gives rise to an
isolated, high viscosity annulus
(Lin et al. 1985)
which attempts to spread with great
urgency into the neighboring regions
of low viscosity.
This spreading leads to a subsequent
depletion of $\Sigma$
in the immediate vicinity of $r_{\rm trigger}$,
and the propagation of two sharp
spikes in $\Sigma$,
one moving to smaller radii
and one to larger radii.
Within these spikes the relative
contrast in $\Sigma(r)$
can be as much as $\sim10^2-10^3$
against the background level.
(One wonders whether some
of our basic assumptions
are  breaking down
  within the spikes.)
The   spikes represent heating
fronts which rush into
the regions of the disk
lying at both smaller and larger
radii
from $r_{\rm trigger}$.
Upon traversing
    a given radial
location,
they initiate a heating from 
the lower to upper S-curve
branch.
Ultimately,
most of the disk transforms to the high state.
It is interesting that regions for
which $\Sigma_{\rm quiescence}(r)$
$ <\Sigma_{\rm max}(r)$
before the arrival of the heating front are 
able to make the cold$\rightarrow$hot transition
quite readily,
due to the large value of $\Sigma$
within the heating front.
The value of $\Sigma(r)$ within the outwardly
progressing heating front
falls as the outer disk edge is approached,
however,
and at some point the condition
$\Sigma_{\rm heating \ front}(r)$
$>\Sigma_{\rm max}(r)$
cannot be maintained.
At this radius the spike in $\Sigma$
associated with the heating front
can no longer
cause heating
because it cannot augment the local, background
$\Sigma$ value to the required level.
The spike stalls and
        collapses down, merging
   back
into the local flow within the disk
from whence it arose.

All of the activity just described
accompanies the observed rise
of an outburst
in a DN or SXT.
   The $\Sigma(r)$ profile
which accompanied the quiescent state 
is radically
transformed 
after the disk finishes
making the low-to-high transition.
The disk attempts to enforce
the condition
${\dot M}(r)$ constant 
with radius as a minimum energy condition.
It has been known for some time that,
for steady or near steady disks
with opacity laws relevant for ionized gas,
$\Sigma(r)\sim r^{-3/4}$.
This is nearly the inverse of the
profile which existed
in the disk during quiescence,
therefore during the subsequent evolution
one sees a juxtaposition
of material from large to small radii.
This activity greatly increases
$\Sigma$ near the inner edge
and thereby increases the level
of accretion onto the central object.
Smak (1984=S84) and Cannizzo et al. (1986=CWP,
see also Cannizzo 1996a)
found that the rise time for outbursts
depends on whether the outbursts
begin at small or   large radii
in the disk $-$
so-called type B or type A outbursts
in Smak's terminology,
or ``inside-out'' versus  ``outside-in''
outbursts in the language of CWP.
The reason for the difference in rise times
versus outburst triggering location
can be understood by looking at the surface
density  evolution.
For type B outbursts,
the surface density at small radii
is small initially,
as is the total surface area of the hot portion
of the disk $A_{\rm hot}$.
The surface density near the central object
sets the level of accretion onto the central object
(Cannizzo 1996b),
and therefore the EUV and soft X-ray fluxes are small.
The optical flux $f_V$
     is set by 
               $A_{\rm hot}$
(Cannizzo 1996b),
and therefore $f_V$ is  also small initially
for the   type B outbursts.
As the outward moving  heating front
travels to  larger radii,
both $A_{\rm hot}$ and $\Sigma(r_{\rm inner})$
increase,
thereby  gradually raising 
         the optical and EUV fluxes.

The behavior for type A outbursts
differs dramatically.
For large $r_{\rm trigger}$ values,
the initial value of $A_{\rm hot}$
just after triggering is
a substantial fraction of the total disk,
which results in
      a rapid increase of the optical flux
to its near-maximum value.
Also,
since most of the mass is stored
at large radii,
this means that the total amount of material
in the inwardly moving heating spike
is also quite large.
   When the inward
traveling heating spike
arrives at the inner edge,
it contains
so much mass that the EUV
flux rises
dramatically
to its maximum value.
One  finds, for SS Cyg
parameters for instance,
that the heating front
arrives at the  inner edge
about one day after the optical flux
rises. Evidence to support this picture
is provided by {\it EUVE} observations
of SS Cyg
     showing an
outburst for which 
  the EUV flux increases by  $\sim10^3$ 
   in a few hours
about a day after the rise
in visual flux
(Mauche 1996).
Mauche also presents observations
of a slow rise outburst in SS Cyg
which is presumably due to an outburst
triggered at a small radius in
the disk.

After the phase of triggering
       and
heating of the  disk has ended
and the $\Sigma(r)$
profile has shifted to $\sim r^{-3/4}$,
there are two possibilities
for the subsequent evolution.
If the surface density
in the outer disk 
 $\Sigma(r_{\rm outer}) < \Sigma_{\rm  min}(r_{\rm outer})$,
and if irradiation of the outer disk is not significant,
then a wave of cooling begins to form at large radii
and starts propagating inward,
causing successive annuli of gas
to make the high-low transition
as it passes.
Just as the heating front operates
by enforcing a local augmentation
of the surface density,
so the cooling wave operates
by enforcing a  local depletion
of the surface density.
This situation comes about because,
at the interface between
the hot,  inner disk and the cool, outer disk
there exists a vigorous outflow of gas.
This outflow is in fact $\sim3 $ times the rate
of accretion onto the central object.
As the matter passes through the transition front,
it cools and its viscosity plummets.
The matter comes almost to a complete halt
(in terms of its radial velocity $v_r$)
once it exits the front at large radii.
By the time the front has traversed the disk
completely to $r_{\rm inner}$,
the entire disk has been transformed back 
into neutral gas.
     The rate of outflow across the front
so greatly   exceeds the rate of accretion
onto the central object
   that
only $\sim2-3$\% of the gas stored
in quiescence is accreted
onto the central object.

If the cooling front is not able to propagate,
either because $\Sigma(r_{\rm outer}) $
$> \Sigma_{\rm min}(r_{\rm outer})$,
or because irradiation is strong enough to keep the outer disk
gas   ionized,
then the disk evolution is much slower.
One must rely on mass loss from the 
inner disk caused by accretion  of gas
onto the central object
to lower the surface density.
After  some time has elapsed
and the disk mass has decreased,
    one has $\Sigma(r_{\rm outer}) < \Sigma_{\rm min}(r_{\rm outer})$,
and/or $T_{\rm irradiation}(r_{\rm outer}) $
$< T_{\rm eff}(\Sigma_{\rm min}(r_{\rm outer}))$
  so that the cooling front can form and begin to 
propagate.
The subsequent evolution is
then as described in the previous paragraph.
One sees evidence of both viscous and thermal decay
in the outbursts of SS Cyg.
The long outbursts
with their flat topped maxima
give evidence for viscous decay
followed by the faster thermal decay,
whereas the short outbursts
which rise to maximum
and then almost  immediately
decay back to  quiescence
show only the thermal decay.
In Cannizzo's (1993b=C93b) model for SS Cyg,
about 3\% of the stored
disk mass is accreted
onto the WD during a short
outburst,
whereas about 30\% is accreted
during a long outburst.





%
%

\section{The ``Problem''  of the UV Delay}


In the computations
of S84
and CWP
which showed evidence
for fast and slow rise
outbursts,
one sees  that the  optical flux
precedes
the FUV flux by $\sim2/3$ day
in the fast rise outbursts.
This is due to the fact that the 
formation of the enlarged
  radiating area
of the disk accompanying 
     the initial triggering
  precedes by $\sim1$ day
the arrival of the heating
spike at the inner disk edge.
A comparison in CWP
of the spacing between
the rise in $V$ and 1050 \AA \ 
flux as seen by the {\it Voyager}
   spacecraft
for a fast rise outburst
in SS Cyg
showed theory and observation to agree.
Both S84 and CWP assumed Planckian
flux distributions
locally for the disk.
Subsequent work by Pringle, Verbunt, \& Wade (1986),
and Cannizzo \& Kenyon (1987)
cast doubt on CWP's findings.
In these later works,
       Kurucz or stellar-like
flux distributions were assumed,
and the delays between rising flux at $V$ and FUV
wavelengths
decreased   dramatically.
The spectral distributions 
           in the later works, however,
also showed sharp absorption edges
due to the Lyman and Balmer series.
La Dous (1989)
compiled a listing of published
Balmer decrements,
most of them measured in SS Cyg
in outburst,
and found them to be smaller
   than expected from theory.
One can understand why the  stellar-like
flux distributions
give such small
FUV delays by comparing stellar  and Planckian
accretion disk flux distributions
for a disk with a given $T_{\rm eff}$
profile (Wade 1984).
In the stellar-like distributions,
the Lyman edge
effectively cuts off the spectrum
below $\sim912$ \AA \ and
redistributes much of this flux into regions
between $\sim1000$ and $\sim2000$ \AA.
Therefore,
if one  compares fluxes at, say,
$\sim1050$ \AA \ and $\sim5500$ \AA \ 
as in CWP,
one finds the $f_{\rm FUV}/f_V$ ratio to be much larger 
for the stellar/Kurucz distributions
than for the  Planckian distributions
(Wade 1984, Fig. 1).
This contributes to an earlier
rise in $f_{\rm FUV}$
with respect to $f_V$
than would have been expected
in the Planckian models.
The earlier works which utilized  the more
primitive Planckian flux distributions
may have characterized better the relevant distributions
inasmuch as the $f_{\rm FUV}/f_V$ ratios were concerned,
because one does not see strong absorption
edges in the observed flux distributions of DNe.
For this reason, I do not believe
there exists a ``UV problem''
as has been repeated so often in the literature.
Until one has in place a reliable
set of basis functions for the accretion disk
flux distributions,
it may be   better to use the light curves
computed from $V$ and ${\dot M}(r_{\rm inner})$
to compare with the observed $V$ and EUV fluxes,
such as those reported in Mauche (1996).

\section{Warner's $M_V$(peak)$-P_{\rm orbital}$
   Relation for DNe and the van Paradijs \& 
  McClintock $M_V-(L_X/L_{\rm Edd})^{1/2}P_{\rm orbital}^{2/3}$
Relation for SXTs}

Warner (1987)
found an empirical relation
between the absolute
visual magnitude
of dwarf novae at maximum light
$M_V$(peak)
and orbital   period $P_{\rm orbital}$.
One can imagine how such a relation might
be expected physically
in the limit cycle model.
As noted earlier,
during the interval  of quiescence
when matter
is being accumulated
as it arrives
from the secondary star,
the $\Sigma(r)$ distribution is bounded
by $\Sigma_{\rm min}(r)$ and $\Sigma_{\rm max}(r)$.
The upper bound $\Sigma_{\rm max}$
ensures that the  disk mass
stored in quiescence cannot exceed a ``maximum mass''
$M_{\rm max}=\int2\pi r dr \Sigma_{\rm max}(r)$.
Now, since $\Sigma_{\rm max}(r)\propto r$,
the maximum mass will be an increasing
function of orbital period.
C93b derived an expression
for $M_{\rm max}(r_{\rm outer})$
by equating the disk mass stored in quiescence
to that in the near steady state disk
which exists just after the  $\Sigma(r)$ profile
has changed from its quiescent shape to its
outburst shape.
The expression obtained by C93b was plotted
by Warner (1995) with the data used in his earlier work,
and there appeared to be consistency.
More recently,
Cannizzo (1998a)
ran time dependent models
for a   range of values  of $r_{\rm outer}$
(equivalent to a range in orbital   period),
to test the accuracy of C93b's analytical estimate.

Does such a relation  exist  for outbursts
in the SXTs?
Van Paradijs \& McClintock (1994=vPMc)
looked at the dependence of
$M_V$
on the soft X-ray flux
$L_X$, expressed in Eddington units,
and $P_{\rm orbital}$.
They 
    determined a
correlation
between $M_V$   and
$(L_X/L_{\rm Edd})^{1/2}P_{\rm orbital}^{2/3}$.
VPMc present a toy model which shows that this
relation is to be expected if the optical flux
is produced
primarily by the irradiation
of the  outer disk
by X-rays coming from accretion
onto the central object.
Indeed,
by comparing the vertical scale
on the $M_V$ plots
presented in Warner (1987) and
vPMc,
     we      immediately
see that SXTs
are $\sim3-7$ mag brighter
at peak outburst than DNe.
For a system at $P_{\rm orbital}\sim8$ hr
representative of our adopted canonical
SXT A0620-00,
we may estimate how large the irradiation effect is.
  The difference  between the two empirically
determined relations at $P_{\rm orbital}\simeq8$ hr
is $\Delta M_V\sim3$ mag.
Some of this difference is due to the 
larger effective area
of the disk around a $\sim7\msun$ BH versus a
$\sim1\msun$ WD,
   at the same  $P_{\rm orbital}$.
For $P_{\rm orbital}$
   constant,
the orbital separation scales
as $M_1^{1/3}$
(if   $M_1>>M_2$),
so the radiating  area of the disk
which gives rise  to the  optical
flux scales as $\sim M_1^{2/3}$.
So in going from a $\sim0.7-1\msun$ WD to 
a $\sim7\msun$ BH, 
     we increase the disk area
by $\sim7^{2/3}-10^{2/3}\sim4-5$,
which reduces $\Delta M_V$
by $\sim2$ mag.
Thus          $\Delta M_V$(corrected) $\sim1$  mag.
During the discussion,  van Paradijs
pointed out that this is only a rough estimate:
there may also be a $\Delta T_{\rm eff}$
correction in going from DNe to BH SXTs.
In any event,
in the next section
we will directly compare
the observed
$M_V$(peak) value determined
from the 1975 outburst of A0620-00
with that  given by  our time dependent
model for A0620-00.
This will provide a more direct
estimate.

\section{The Ubiquitous Exponential Decays
   in DNe and SXTs}

Bailey (1975) looked at the
decay light curves
of DNe spanning
a range in   orbital   period
and determined
the decays to be roughly exponential
  $-$ the associated decay time constant
scaling with orbital period.
The ``Bailey relation'' is 
$t_e(V)\simeq3$ days 
   [$P_{\rm orbital}/P_{\rm orbital}$(SS Cyg)].
It is interesting to note that,
for many of the brightest and
best studied SXTs,
including A0620-00,
one also finds exponential decays.
The associated $e-$folding  time constant
is roughly a factor of 10 slower
than that of DN decays.
This can be seen
directly in the famous plot of
four BH SXT
outburst light curves
which has appeared in so many review
articles
  (cf.
Tanaka \& Shibazaki 1996, Fig. 3).
In general,
all the time scales
associated with
outbursts in the SXTs
are much longer than those
associated with outbursts in the DNe.
This seems to be true
for both the NS SXTs and the BH SXTs,
therefore arguing for an irradiation
effect and not a scaling with 
central object mass
to account for the differences $-$
assuming the basic limit cycle model
to be at work in both systems.
Another possibility,
which I will argue for below,
is that irradiation accounts for the 
difference in going from DNe to NS SXTs,
but differences in central object mass
$M_1$ account for differences in
the light curves in going from DNe to BH SXTs.

In his talk,
Andrew King argued that the irradiation
effect is strong for the BH SXTs
in outburst  (cf. King \& Ritter 1997=KR).
The irradiation flux scales as $r^{-2}$,
whereas the viscous flux scales as
$r^{-3}$ for a steady state disk.
So, all other factors being equal,
one would indeed expect irradiation
to become important
at large radii where the bulk of the
optical flux is produced.
KR argue
that irradiation prevents the cooling
front from forming and thereby
keeps the entire disk in the hot,
ionized state.
Cannizzo (1994) pointed out that,
if the limit cycle model is operating
and the disk is in outburst,
the profile of aspect ratio $h/r$ versus $r$
will acquire a convex shape.
This would shield 
        the outer disk from 
direct irradiation.
It is not clear, however,
how reliable the theoretical
$h/r$ values are for actual disks.
Also, indirect irradiation from, for example,
a scattering corona above the disk,
could alter this
simplistic picture
of the shielding.

The general scenario of the viscous decay
advocated by KR has been studied
in detail by many workers
(e.g. 
   Bath, Clarke, 
  \& Mantle 1986,
Lyubarskii \& Shakura 1987, 
    Cannizzo, Lee, \& Goodman 1990=CLG,
Mineshige, Yamasaki,
\& Ishizaka 1993).
In view of the complex  calculations
required to model the limit cycle
instability in accretion disks,
it is interesting to note that
the viscous decay process $-$ 
in which the disk mass
can only decrease by virtue
of mass loss   onto the central object $-$
is remarkably simple:
the evolution can be expressed analytically.
CLG present a convenient
   power law  
     form
for the decay of the disk mass
$dM_{\rm disk}/dt\sim t^{-q}$,
where $q=(38+18a-4b)/(32+17a-2b)$
and the Rosseland mean opacity
$\kappa=\kappa_0\rho^a T^b$.
This scaling assumes $\alpha$ to be constant.
The   parameter   $q$  is relatively insensitive
to the opacity:
for electron scattering  $(a=0,b=0)$
we get $q=19/16$,
whereas for Kramer's opacity $(a=1,b=-3.5)$
we get $q=5/4$. 
If one plots
this functional form
for the decay on the
same scale as the 1975 outburst
of A0620-00
as seen in X-rays,
one sees that it does not at all
resemble the observed decay.
The reason for this failure
can be seen trivially
in the structure of the diffusion 
equation for surface density,
$\partial\Sigma/\partial t\propto $ fcn($\nu\Sigma$),
where $\nu\propto\alpha T_{\rm midplane}$.
For $\alpha$ constant,
one typically has $T_{\rm midplane}\propto \Sigma^{2/3}$
(this is exact if the opacity is
  electron scattering),
so that     $\partial\Sigma/\partial t\propto\Sigma^{5/3}$.
For exponential decays
one needs   $\partial\Sigma/\partial t\propto\Sigma$
so that $\Sigma(t)=\Sigma_0 e^{-t/t_0}$.
One can ensure exponential
decays by having $\nu$ be constant,
therefore requiring either an isothermal disk
if $\alpha$ is constant,
or $\alpha\propto 1/T_{\rm midplane}$.
This scaling seems unphysical
and would have rather
dire consequences
for the limit cycle picture.

What are the equivalent constraints
imposed on $\alpha$
by starting with the assumption
that the exponential decays
are due to a cooling wave
traversing the disk
and shutting off the flow
onto the central object?
Vishniac \& Wheeler (1996=VW,
   see also Vishniac 1997)
present an analytical
theory for the functional form
of the decay of the light curve
in an outburst caused
by the limit cycle instability.
Their model relies on deviations
from steady state flow conditions
within the inner, hot portion
of the disk during the decay from
maximum light.
VW's theory makes some detailed predictions
about the flow patterns within the hot
   portion of the disk
which
are substantially
borne out in detailed testing
(Cannizzo 1998b).
One may use their analytical formalism
to derive an expression for
the rate of decay of the [hot] disk mass
for $\alpha$ constant.
If one plots
this functional form
for the decay on the
same scale as the 1975 outburst
of A0620-00
as seen in X-rays,
one sees that it does not at all
resemble the observed decay.
The reason for  this failure
can be understood
by  considering the scaling
of the cooling front velocity
with time.
For $\alpha$ constant,
the cooling front has a faster-than-exponential
speed (when expressed as a function   of
radial position)
leading to a  faster-than-exponential
decay of the light curve.
This is in contrast with the  purely
viscous decay model,
in which $\alpha=$ constant
leads to a power law
(slow-than-exponential) decay.
Therefore, to get the cooling wave
model to work,
      we have just
the opposite problem as with the
viscous decay model:
now we need to have $\alpha$
decrease as the outburst proceeds.
Cannizzo et al. (1995=CCL) tested functional forms
for $\alpha$ and concluded
   that
the exponential decays
seen in the BH SXTs
could be produced
with a cooling wave provided
$\alpha=\alpha_0(h/r)^n$,
where $\alpha_0\simeq50$ and $n\simeq1.5$.
This form for $\alpha$ was first used
by Meyer \& Meyer-Hofmeister (1983, 1984)
in their modeling of DN outbursts,
although they did not justify
it in detail observationally.
In CCL's model,
the $e-$folding decay time
    $|d\ln M_{\rm disk}/dt|^{-1}$ is given
by $t_e\simeq 0.4GM_1\alpha_0^{-1} c_s^{-3}$,
where $c_s$ is a constant
$\sim16$ km s$^{-1}$.
(As a minor aside,
the value $\alpha_0\simeq50$
presupposes $M_1\sim10\msun$
   in the BH SXTs.
A recent compilation of inferred
SXT BH masses by Bailyn et al.
(1996) shows a strong peak in the distribution
at $\sim7\msun$.
If true, this would lower $\alpha_0$ to
$\sim35$.)
The scaling $t_e\propto M_1$
would account naturally
for the  factor
of $\sim10$
difference between DNe and BH SXTs.
It would not explain NS SXTs,
where irradiation appears to be
an important factor
in determining the outburst
characteristics
(van Paradijs 1996).

It   is interesting to note
that neither
the viscous model
nor the  cooling wave
model for the decays of outbursts
naturally produces an exponential
decay if $\alpha$ is constant.
One must assume that one of the models
is correct,
and then infer the constraint on $\alpha$
required to get the model to work.
That fact that $\alpha=$ constant
fails fits in with the theme of the talk
by John Hawley.
He stressed that we only
use ``$\alpha$'' 
to enable standardized comparison of
results between different groups of workers,
   and that the mere  usage of
   ``$\alpha$'' as a concept
   can engender a false 
   sense of understanding
  or foster
   predisposed  notions
   about the  physics of viscous
dissipation  and angular momentum
   transport
and their (untenable)
  connection  with turbulence.
The physical dependence
of $\alpha$ on local or global physical
conditions within the  disk
may require many years of  MHD modeling
to ascertain.
Such efforts to date
have only gone as
far as
determining
$\alpha\simeq0.01-0.1$,
generally consistent
with values inferred
from studies of DN and SXT
outbursts.
As a further note,
we stress
that the aforementioned
constraints
   derived from the exponential decays
only apply to 
$\alpha_{\rm hot}$
   which is pertinent to the
    ionized gas
on the upper stable branch
of the S-curve,
for it is the hot part of the
accretion disks
that give rise to the optical
and soft X-ray fluxes we observe.
The viscosity parameter $\alpha_{\rm cold}$
along the lower branch of the S-curve
of relevance to the neutral gas
is set primarily by the
recurrence
time scales for outbursts
   (Cannizzo, Shafter, \& Wheeler 1988).
It may prove difficult
to constrain a functional 
form for $\alpha_{\rm cold}$.

There is one additional
consideration
involving
the application
of the law $\alpha\simeq50(h/r)^{1.5}$
to DNe.
CCL showed,
by varying the outer radius
of the disk
in a series of models,
that the $e-$folding
time for outbursts
is relatively
insensitive
to $r_{\rm outer}$
and therefore $P_{\rm orbital}$.
This result stands in contrast
to previous disk instability computations
which  showed general  consistency
between theory and observation
as regards the Bailey relation
(S84, Cannizzo 1994).
A resolution of this contradiction
may   lie within
recent work by Gu \& Vishniac (1998),
who redo
the vertical structure
computations
(which give us
    the  S-curve)
taking $\alpha=50(h/r)^{1.5}$.
Rather than prescribe
$\alpha$ to be constant
as in previous works,
Gu \& Vishniac
     iterate  to get $\alpha$
for each value of ${\dot M}$
in a series of models
at fixed $r$ and $M_1$.
They find  that,
due to an inherent inverse scaling
of $h/r$
with $M_1$,
the $\alpha$ values along the upper branch
of the S-curve tend to be larger for $M_1=1\msun$
than for $M_1=10\msun$.
In fact, $\alpha_{\rm hot}$
exceeds the typical values
which S84 found are required to get the Bailey relation.
Therefore $50(h/r)^{1.5}$
cannot be the entire scaling law:
whatever physical mechanism produces this
scaling (if it is in fact correct)
must saturate at some $(h/r)_{\rm crit}$
so that $\alpha_{\rm crit}\sim 0.1-0.2$.
The  existence of a saturation value to $\alpha$
would restore the Bailey relation for DN parameters.

While we have many
convincing examples
of quantifiably
exponential decays
in the DNe and SXTs,
do we have any good examples
of power law decay?
As mentioned earlier,
the flat topped
outbursts
in SS Cyg
are supposedly viscous
rather than thermal,
but unfortunately
the dynamic range
in flux level
of a flat topped
outburst
is not large
enough to reveal
what the functional form of the decay is.
Only the WZ Sge stars
(including WZ Sge itself)
have enough dynamic range
during their superoutbursts
to be
able to see the
slower-than-exponential
form
characteristic of viscous decay,
a feature also seen in
models for superoutbursts
(Osaki 1996).
The only other example I have been
able to find
is V1057 Cyg (Cannizzo 1996a),
an FU Ori star.

\section{Long Term Light Curves of the SXTs}

The $\alpha$ value  inferred
by CCL was based solely on the 
decay properties
of the BH SXTs.
The obvious next step
is to run models covering
complete cycles of outburst
and quiescence for parameters
appropriate to A0620-00
to see
if the outbursts
bear any resemblance
to those observed.
Two outbursts
have been seen in A0620-00
(Nova Mon 1917 and 1975).
Cannizzo  (1998b)
performs long runs
using the time dependent
model described in CCL
and finds that the outburst
properties are quite different
from those
observed.
Although the decays of the outbursts
have the $\sim30$ day
$e-$folding time scale,
the outbursts
occur every $\sim5$ yrs
and have rise times much slower than
those observed.
Most of the outbursts are also
low amplitude,
${\dot M}_{\rm peak}\sim10^{15}-10^{17}$ g s$^{-1}$,
in contrast to the  value $\sim10^{18}$ g s$^{-1}$
inferred for A0620-00.
   The 1975 outburst
of A0620-00 made it a sustained
     $\sim50$ Crab source
for $\sim2$ wks
    in $3-6$ keV X-rays
as observed by {\it Ariel 5}.
Much fainter
 outbursts,
had they existed,
would have  been observed with a
host  of X-ray  satellites
over the past $\sim20$ yrs.
Therefore
one must conclude that the standard
model
does not work very well.

It has   been known for some time that
the quiescent
observations
of DNe and SXTs
are not consistent with theory.
One sees
substantial EUV and soft X-ray  fluxes
in quiescence,
and yet in the standard limit cycle disk
instability model
one expects negligibly small rates
of accretion  onto the  compact object.
Some mechanism extrinsic
to the standard model  must  evaporate
or remove    gas  from the inner
disk and deposit it onto  the compact
object  in such a fashion
that it does not alter
the quiescent $\Sigma(r)$ profiles
to the extent that an outburst is triggered.
Meyer \& Meyer-Hofmeister (1994)
proposed
that a hot, ``coronal siphon flow''
exists in the  inner disk.
Hot electrons  in a corona conduct heat
down to the disk photospheric layers
and drive a flow
onto the
central object.
Liu, Meyer, \& Meyer-Hofmeister
(1995)
     show general  consistency
between theory 
and   observation for certain DNe.
While the exact physics discussed by
  Liu et al.
may or may not   be   relevant,
it is clear that,
in order to have a complete
theory   for the light   curves
of DNe and BH SXTs
covering not only outbursts
but also quiescence,
some evaporative process
must   be   at work.
Therefore
for full generality
we should include the effects
of evaporation
or mass removal  from the
inner disk   in our long term computations.

C97b  finds that when evaporation
is included,
the frequent, small amplitude
outbursts
which are triggering at small radii
in the standard model
are eliminated.
One is only left  with major  outbursts
(${\dot M}_{\rm peak}\sim10^{18}$ g s$^{-1}$)
occurring every $\sim50-100$ yrs,
as observed.
By  eliminating the inner disk,
the rise times of the  outbursts 
are also  shortened
somewhat because the triggering
must occur further out in the disk.
(We move from having  type B to type A outbursts.)
The rise times are still slower than
observed, however.
Recent time dependent disk instability
computations for the  galactic
microquasar GRS 1915+105
(Hameury et al. 1997)
find that,
for parameters
relevant to this  long orbital period/
high mass transfer rate system,
the rise times are significantly
affected
by the altered $\Sigma_{\rm quiescence}(r)$
profile
produced
by introducing evaporation.
It is not clear why our modeling
of A0620-00 differs
in this respect.

The   observed 1975 outburst
of A0620-00
had a peak visual magnitude
corrected for extinction
$M_V$(peak)$=+0.7$ (vPMc),
a total energy $\delta E=3\times10^{44}$
ergs
(Chen, Shrader, \& Livio 1997)
which gives an accreted mass
$\delta M\sim5\times10^{24}$ g
(assuming a rest mass to energy  conversion
efficiency $\epsilon\sim0.06$),
and a peak rate of accretion
within the disk
$\sim10^{18}$ g s$^{-1}$.
These values are all close to those 
    found
by C97b
using the standard
model plus evaporation from the inner disk at
a rate ${\dot M}_{\rm evap}\sim10^{14}$ g s$^{-1}$.
We may estimate how strong irradiation
is at the peak
of the outburst
by directly comparing
observation with theory.
This should provide
a more  specific estimate for A0620-00
than was given in the previous section.
Our computed $M_V\simeq +1$   for a face-on disk.
Taking a $60^{\circ}$
inclination
thought to be appropriate for A0620-00
reduces
the optical flux by $\sim\cos 60^{\circ}=0.5$
which adds
$\sim1$ mag to $M_V$(peak).
This difference
is similar to that found
by previous workers
who modeled
     $B$ and $V$ for 
  the  1975
outburst of A0620-00
using   nonirradiated
time dependent models for the
accretion disk limit cycle,
and compared their
flux levels
directly with 
A0620-00
(Huang \& Wheeler 1989,
Mineshige \& Wheeler 1989).
Therefore $\Delta M_V$
between observation and theory (without irradiation
included) is $\sim1$ mag,
in line with our previous estimate  obtained
by comparing the Warner and vPMc
relations.
As a final check,
C97b also runs models
with irradiation included
(as described
  in Mineshige, Tuchman, \& Wheeler 1990)
and  finds that,
within the inner hot portion
of the disk existing
at the time corresponding to
maximum optical light
in the outburst,
the ratio   of the local irradiation
temperature to the local
(viscous)
effective temperature
is $\sim0.3-0.4$.
Therefore,
were we to   recompute
the visual  magnitude
with irradiation
included,
this would  increase the effective
temperature
in the  outer portion of the hot,
inner disk by $\sim1.3-1.4$.
    Since the visual flux scales
as $\sim T_{\rm eff}^2$
for the temperature region of interest
(vPMc),
this would  increase
the optical flux by $\sim2$
and account for the  $\Delta M_V\sim1$ discrepancy
between   the observed
value of  $M_V$(peak)
and the value  taken from  the nonirradiated  models.
If most of the area of the entire disk
(going out to $\sim0.7$ of the Roche lobe of 
the primary) were strongly
irradiated as in KR 
so that the cooling  front  was
prevented from forming
and   propagating,
then the irradiation-induced
optical flux would far exceed that observed.

The model of C97b which  includes
the standard
model plus evaporation
for A0620-00
has ${\dot M}_{\rm evap}\sim10^{14}$ g s$^{-1}$,
and yet the value inferred
by McClintock, Horne, \& Remillard (1995)
for the quiescent
value of ${\dot M}$
in A0620-00 inferred
from {\it  ROSAT}
observations
is $\sim10^{10}-10^{11}$ g s$^{-1}$.
This assumes an accretion   efficiency $\epsilon\sim0.1$.
The discrepancy might be accounted for in the models 
of Narayan, McClintock, \& Yi (1996) 
   which assume a much smaller efficiency
$\epsilon\sim10^{-4}$
for the quiescent state.
On the other hand,
if the model of Narayan et al. were correct,
one would expect 
     a large difference $\sim10^3$
between
the quiescent  flux  level of soft X-rays
in going from
the BH SXTs
to the NS SXTs
because, in the NS SXTs
the infalling material strikes
the neutron star surface
and thus $\epsilon\sim0.1$,
whereas for the BH SXTs
most of the thermal energy
is supposedly
advected 
     through the event horizon.
The observed quiescent
fluxes in the NS and BH
systems
are actually quite similar,
however
 (Chen, Shrader, \& Livio 1997),
despite claims to the 
contrary (Narayan, Garcia,
  \&  
  McClintock 1997).
Although  evaporation is attractive
in some aspects,
it appears to be problematic  in
other ways.
C97b discusses other possible resolutions
to the failure of the standard
model (i.e., without resorting to
evaporation)
in the BH SXTs.

\section{Conclusion}

We present
an overview
of accretion disks in transient
systems,
with concentration
on the limit cycle
instability model
which has been
developed extensively
to account
for the outbursts
seen in DNe and SXTs.
In much of our discussion
we use the two brightest
members
of these classes
in our time dependent modeling $-$
SS Cyg for DNe and A0620-00 for SXTs.
Our conclusions  are  drawn primarily
from comparing  theory  with observation
for these two systems.
Since A0620-00  is a black hole candidate,
our conclusions regarding the BH SXTs
have been developed in more detail 
than for the NS SXTs.

Our conclusions are as follows:

(i)  If the instability triggers near
the inner disk edge
it   produces a slow rise outburst;
if it triggers some distance away
from the inner edge  it produces
a fast rise outburst.
The differences are due
to the rate of build-up
in surface density
near the compact object,
  and the rate of enlargement
of a hot area in the disk
which produces the optical flux.

(ii) Until ``correct'' flux distributions
are used
for the  accretion  disk,
one cannot determine whether or not there
exists
a problem with the model
as regards
the delay of the FUV
(i.e., flux from $\sim1000-2000$
\AA)  with respect
to the $V$ during a fast rise outburst.

(iii) For both DNe and SXTs
there exists
a relation
between
$M_V$(peak)
and
systemic
parameters
$-$
$P_{\rm orbital}$
for the DNe
and $(L_X/L_{\rm Edd})^{1/2} P_{\rm orbital}^{2/3}$
for the SXTs.
The absolute level of $M_V$(peak)
between the two classes
indicates
brighter disks
in the SXTs
which is due in part
to irradiation
and in part to larger
disks.
The underpinnings of the 
Warner  relation
derive    from the concept of the ``maximum
mass''             which can be stored
in the quiescent state
   before the  next outburst
      is triggered
   (Cannizzo, Shafter, \& Wheeler 1988).

(iv)  The exponential decay in soft X-ray
flux of outbursts
seen in SXTs
places
constraints
on the viscosity
parameter $\alpha$.
For constant $\alpha$,
both the viscous decay and cooling
wave models fail:
the former produces a slower-than-exponential
decline in $dM_{\rm disk}/dt$,
whereas the latter
produces
a faster-than-exponential decay.
To avoid these unpleasant  attributes
and obtain agreement with observations,
one would require $\alpha$
to increase with time as the outburst decays
for the viscous decay model,
and to decrease with time as the outburst
decays in the cooling wave model.
 A calculation of $\alpha$ from first
principles would  help decide
which option is more physically
plausible.

(v) In the cooling wave model for the decay,
   one requires $\alpha=\alpha_0(h/r)^n$,
  where $\alpha_0\sim35-50$ and $n\simeq1.5$.
For $n=1.5$
the $e-$folding decay  time  for the mass
contained
within the inner,
hot portion
of the disk
is
$t_e\simeq0.4 GM_1\alpha_0^{-1}c_s^{-3}$
$\propto M_1$,
where $c_s\simeq16$ km s$^{-1}$.
The fact that the $e-$folding decay time
of a system like A0620-00
is $\sim30$ days
whereas
that for a system like SS Cyg
is $\sim3$ days
would be a consequence
of the ratio $M_1$(BH)/$M_1$(WD).
The outbursts in the NS SXTs
(where $M_1\sim1\msun$
as for DNe)
are complicated
by the   effects
of irradiation
and would not presumably follow this law.

(vi) The ``standard model''
(i.e., $\alpha=50(h/r)^{1.5}$)
of SXTs
produces major outbursts
with recurrence times
and ${\dot M}_{\rm peak}$
values similar to those observed,
but one also sees
frequent,
low amplitude
outbursts
which   were not seen in A0620-00.
The introduction of evaporation
of matter
from the 
inner disk gets rid
of the small
outbursts
and leaves
only the major ones,
but does not reduce the 
rise times
of the major outbursts
to a value
as small
as that seen in the 1975
outburst
of A0620-00.
This is the only shortcoming of our A0620-00
model and may prove to be minor.
Alternatively,
   another factor
such as a lower
   $\alpha_0$ value
   in quiescence
   may suppress
  type B outbursts.
Hameury et al. (1997)
find better agreement
between theory and  observation
as regards the rise time
characteristics
at different 
wavelengths
using evaporation
in disk instability computations
for GRS 1915+105.

(vii) By comparing
theory and observation
for the value of $M_V$
at maximum light in A0620-00,
we find a shortfall of $\sim1$ mag
in the theory which seems
plausibly to be accounted for
by introducing
self-irradiation
of the disk
into the model,
in agreement with previous workers.

\acknowledgments

I thank Wan Chen, Mario Livio,
  Andrew King, Jan van Paradijs,
Brian Warner, Chris Mauche,
Ethan Vishniac, Craig Wheeler, 
and Jim Pringle for useful discussions.
  JKC
 was supported 
through the long-term
scientist program under
 the Universities Space Research Association
(USRA contract NAS5-32484)
in the Laboratory for High Energy Astrophysics
at Goddard Space Flight Center.

\end{document}